%% file: main.tex
\documentclass[conference]{IEEEtran}
\IEEEoverridecommandlockouts
\usepackage{blindtext, graphicx}
\usepackage{cite}
\usepackage{microtype}
\usepackage[cmex10]{amsmath}
\usepackage{amssymb}
\usepackage{amsfonts}
\usepackage{mathtools}
\DeclarePairedDelimiter{\ceil}{\lceil}{\rceil}
\usepackage{algorithm}
\usepackage[noend]{algpseudocode}
\usepackage{url}
\usepackage{todonotes}
\usepackage{textcomp}
\usepackage{parskip}
\usepackage{pgfplots}
\pgfplotsset{compat=newest}
\usepgfplotslibrary{colorbrewer}
\pgfplotsset{cycle list/Dark2-8}
\usepackage{pgfplotstable}
\usepackage{booktabs}
\usepackage{tabularx}
\usepackage{multirow}
\usepackage{tablefootnote}
\usepackage{footmisc}
\usepackage{comment}


\usepackage[pdftex,hidelinks]{hyperref}

\definecolor{LHS}{RGB}{0, 80, 158} 
\definecolor{RHS}{RGB}{92, 190, 201} 

\newcommand{\OurScheme}{\textsc{BISMO}}
\graphicspath{{figures/}}
\DeclareGraphicsExtensions{.pdf}
\pgfplotsset{compat = 1.3}

\begin{document}

\title{\OurScheme{}: A Scalable Bit-Serial \\ Matrix Multiplication
  Overlay for \\ Reconfigurable Computing
}

\author{\IEEEauthorblockN{Yaman Umuroglu$^\dagger$}
\thanks{$^\dagger$Corresponding author. Work performed while author was at NTNU.}
\IEEEauthorblockA{
yamanu@xilinx.com \\
Xilinx Research Labs \\
Dublin, Ireland}
\and
\IEEEauthorblockN{Lahiru Rasnayake and Magnus Sj\"alander}
\IEEEauthorblockA{
[firstname.lastname]@ntnu.no \\
Department of Computer Science\\
Norwegian University of Science and Technology (NTNU)\\
Trondheim, Norway}
}

\maketitle

\input{content/0-abstract}


%
\IEEEpeerreviewmaketitle

\input{content/1-introduction}
\input{content/2-background}
\input{content/3-scheme}

\input{content/4-results}
\input{content/5-relatedwork}
\input{content/6-conclusion}

\bibliographystyle{ieeetr}
\bibliography{defs-short,refs}

\end{document}

%% file: content/0-abstract.tex
\begin{abstract}
  Matrix-matrix multiplication is a key computational kernel for
  numerous applications in science and engineering, with ample
  parallelism and data locality that lends itself well to
  high-performance implementations. %
  Many matrix multiplication-dependent applications can use
  reduced-precision integer or fixed-point representations to increase
  their performance and energy efficiency while still offering
  adequate quality of results. %
  However, precision requirements may vary between different application phases
  or depend on input data, rendering constant-precision solutions
  ineffective.  %
  We present \OurScheme{}, a vectorized bit-serial matrix
  multiplication overlay for reconfigurable computing. %
  \OurScheme{} utilizes the excellent binary-operation performance
  of FPGAs to offer a matrix multiplication performance that scales
  with required precision and parallelism. %
  We characterize the resource usage and performance of \OurScheme{}
  across a range of parameters to build a hardware cost model, and
  demonstrate a peak performance of 6.5~TOPS
  on the Xilinx PYNQ-Z1 board.
\end{abstract}


%% file: content/1-introduction.tex
\section{Introduction}

Using constant precision for all operations is the predominant practice when
designing digital systems, since logical and arithmetic operations, registers,
memories, and interconnects can be designed to accommodate one specific
precision. %
Their main disadvantage is the associated overhead in storing, communicating,
and performing operations with full precision when an application only requires
a fraction of the supported precision. %
Numerous applications, in the engineering, scientific, and multimedia
domain, can use reduced precision and still produce adequate results. %
This property has been leveraged in approximate
computing~\cite{mittal+:CSUR2016approx_survey} and quantized neural
networks (QNNs)~\cite{hubara2016quantized, park2017weighted},
to improve performance and energy efficiency and to reduce
area by tailoring computations to the required precision. %
This precision may vary between different phases of the application.
As an example, Park et al.~\cite{park2017weighted} achieve the best
performance-accuracy tradeoff for QNNs by using fewer bits for the intermediate
layers.

Matrix-matrix multiplication is a commonly used computational kernel and
represents one of the seven Berkeley dwarfs, which are
important computational constructs for engineering and scientific
computing~\cite{asanovic+:dwarfs-2006}. %
The amount of computations required for matrix multiplications makes it highly
beneficial to adapt the operational precision to an application's
requirements. %
FPGAs are a good fit for low-precision operations and
for instantiating efficient matrix multiplication accelerators with a specific
precision. 
However, fixed-precision accelerators are not suitable for applications with
variable precision as they either
require multiple instances of the same accelerator, each with a different precision,
or require dynamic reconfiguration with associated overhead and system
complexity. 

A promising alternative to fixed-precision accelerators is to use bit-serial
computations~\cite{umuroglu_jahre:CASES2017} where the integer matrix
multiplication is expressed as a weighted sum of binary matrix multiplications
(\autoref{sec:background}). %
The bit-serial alternative provides the possibility to use an efficient binary
matrix multiplication accelerator to compute matrix multiplications of any
precision. %

We present a scalable bit-serial matrix multiplication overlay called
\OurScheme{} that can be efficiently instantiated on an FPGA. %
The core of \OurScheme{} is a software-programmable weighted binary
matrix multiplication engine and associated hardware for fetching data and
storing back the result (\autoref{sec:hw_architecture}). %
The hardware architecture is design-time configurable and we provide a cost
model for estimating the resource usage for a given set of parameters
(\autoref{sec:cost_model}).
\OurScheme{}'s software programmability makes it possible to operate on any
matrix size and any fixed-point or integer precision (\autoref{sec:sw_stack}). %

We evaluate \OurScheme{} on the Xilinx PYNQ-Z1 board and show %
\textit{i)}~that the number of look-up-tables (LUTs) scales linearly with the
number of parallel binary dot-product operations and
\textit{ii)}~an average 94\% accuracy for the proposed cost model
(\autoref{sec:synthesis}). %
We also show that the performance scales linearly with allocated resources and
that the runtime scales better than expected when increasing the dot-product
precision. %
\OurScheme{} achieves a peak performance of 6.5 binary TOPS and an energy
efficiency of up to 1.4 binary TOPS/W (\autoref{sec:performance}), which is
best-in-class with only a dedicated ASIC accelerator showing better performance
(\autoref{sec:relatedwork}). %
\OurScheme{} is open-sourced at https://git.io/fWb0m.


%% file: content/2-background.tex
\section{Bit-Serial Matrix Multiplication}
\label{sec:background}


Fixed-precision operations have to be designed to accommodate the
largest supported precision, which causes overheads in cases where the
required precision of an application varies throughout its execution
or when the precision depends on its input data. %
In contrast, bit-serial operations are inherently frugal since they
only compute as many bits as specified by the precision of the
operands. %
However, their serial nature causes high latencies and potentially poor
performance. %

Matrix multiplication is a suitable kernel for taking advantage of the
frugality of bit-serial operations while overcoming the high-latency
by performing many bit-serial operations in parallel. %
Umuroglu and Jahre showed that by expressing a matrix multiplication
as a weighted sum of binary matrix multiplications
(Algorithm~\ref{alg:bit-serial_matrix_multiplication}) it is possible
to efficiently compute matrix multiplications of variable precision
using the logical \textsc{And} and population count (popcount)
instructions available in most modern
processors~\cite{umuroglu_jahre:CASES2017}. %
In addition, the algorithm works for both integer as well as fixed point
number representations, where the new fixed point location is given by the
product of the input matrices' scaling factors. %

\begin{algorithm}[t]
  \small
    \centering
    \begin{algorithmic}[1]
      \algrenewcommand\algorithmicindent{1.0em}
      \State \textbf{Input:} $m \times k$ $l$-bit matrix $L$, $k \times n$ $r$-bit matrix $R$
      \State \textbf{Output:} $P = L \cdot R$
      \For{$i \gets 0 \dots l-1$}
        \For{$j \gets 0 \dots p-1$}
          \State $\mathrm{sgnL} \gets (i == l-1 \mathrel{?} -1 : 1)$
          \State $\mathrm{sgnR} \gets (j == p-1 \mathrel{?} -1 : 1)$
          \State $\mathrm{weight} = \mathrm{sgnL} \cdot \mathrm{sgnR} \cdot
          2^{i+j}$
          \State \textit{\#  Binary matrix multiplication between
            $L^{[i]}$ and $R^{[j]}$ }
          \For{$r \gets 1 \dots m$}
            \For{$c \gets 1 \dots n$}
                \For{$d \gets 1 \dots k$}
                    \State $P_{rc} = P_{rc} +  \mathrm{weight} \cdot (L^{[i]}_{rd} \cdot R^{[j]}_{dc}) $
                \EndFor
            \EndFor
        \EndFor
        \EndFor
      \EndFor
    \end{algorithmic}
    \caption{\small Bit-serial matrix multiplication on signed integers.}
    \label{alg:bit-serial_matrix_multiplication}
\end{algorithm}


\autoref{fig:bit-serial_example} illustrates
Algorithm~\ref{alg:bit-serial_matrix_multiplication} for the example
where the two input-matrices ($L$ and $R$) consists of 2-bit unsigned
integer numbers. %
By expressing $L$ and $R$ as weighted sums of binary matrices, the
matrix product ($L\cdot R$) can be expressed as a weighted sum of
products between binary matrices. %
The matrix multiplication can thus be expressed as a large number of
binary operations that can be performed in parallel. %

\begin{figure}[th]
\begin{equation*}
\begin{aligned}
  L = &
  \begin{bmatrix}
    2 & 0 \\
    1 & 3
  \end{bmatrix}
  = 2^1 \textcolor{blue}{L^{[1]}} + 2^0 \textcolor{red}{L^{[0]}}
  = 2^1
  {
  \color{blue}
  \begin{bmatrix}
    1 & 0 \\
    0 & 1
  \end{bmatrix}
  }
  + 2^0
  {
  \color{red}
  \begin{bmatrix}
    0 & 0 \\
    1 & 1
  \end{bmatrix}
  }
  \\
  R = &
  \begin{bmatrix}
    0 & 1 \\
    1 & 2
  \end{bmatrix}
  = 2^1 \textcolor{blue}{R^{[1]}} + 2^0 \textcolor{red}{R^{[0]}}
  = 2^1
  {
  \color{blue}
  \begin{bmatrix}
    0 & 0 \\
    0 & 1
  \end{bmatrix}
  }
  + 2^0
  {
  \color{red}
  \begin{bmatrix}
    0 & 1 \\
    1 & 0
  \end{bmatrix}
  }
  \\
  P = & 
  L \cdot R = 
  (2^1 \textcolor{blue}{L^{[1]}} + 2^0 \textcolor{red}{L^{[0]}})
  \cdot
  (2^1 \textcolor{blue}{R^{[1]}} + 2^0 \textcolor{red}{R^{[0]}}) \\
  = &
  2^2 \textcolor{blue}{L^{[1]}} \cdot \textcolor{blue}{R^{[1]}}
  +
  2^1 \textcolor{blue}{L^{[1]}} \cdot \textcolor{red}{R^{[0]}} +
  2^1 \textcolor{red}{L^{[0]}} \cdot \textcolor{blue}{R^{[1]}}
  +
  2^0 \textcolor{red}{L^{[0]}} \cdot \textcolor{red}{R^{[0]}} \\
\end{aligned}
\end{equation*}
\caption{Example of a bit-serial matrix multiplication on unsigned
  integers with the two first for-loops unrolled
  (Algorithm~\ref{alg:bit-serial_matrix_multiplication}: for-loop on
  line 4 and 5 unrolled and weight on line 8 always positive).}
\label{fig:bit-serial_example}
\end{figure}


%% file: content/3-scheme.tex
\section{The Bit-Serial Matrix Multiplication Overlay}


\OurScheme{} consists of a hardware part and a software part. %
The hardware part is composed of a scalable bit-serial matrix
multiplication datapath and associated memory and control logic. %
The software part generates instructions for the hardware for a given
matrix size and precision. %
The key features offered by this hardware-software design are as
follows: %

\textbf{Precision-scalable.} %
By expressing an integer or fixed-point matrix multiplication as a
weighted sum of binary matrix multiplications
(\autoref{sec:background}), the same hardware can be utilized for a
range of different precisions. %
Lower-precision matrix multiplications are finished quickly, while
higher-precision requires more clock cycles. %

\textbf{Hardware-scalable.} %
Our overlay generator can scale the memory and compute resource
utilization to match system-level requirements. %
This is achieved by controlling the parameters described in
\autoref{sec:hw_architecture}. %
We also provide a cost model to estimate the resource usage for a
given set of parameters as described in \autoref{sec:cost_model}. %

\textbf{Software-programmable.} %
Our hardware architecture is software-programmable at the granularity of
instructions as described in \autoref{sec:sw_stack}. %
This offers several advantages such as the ability to tailor block sizes and
dynamically skip bit positions for sparse or approximate computing. %

\subsection{Hardware Architecture}
\label{sec:hw_architecture}

\begin{figure}[t]
  \centering
  \includegraphics[width=0.96\columnwidth]{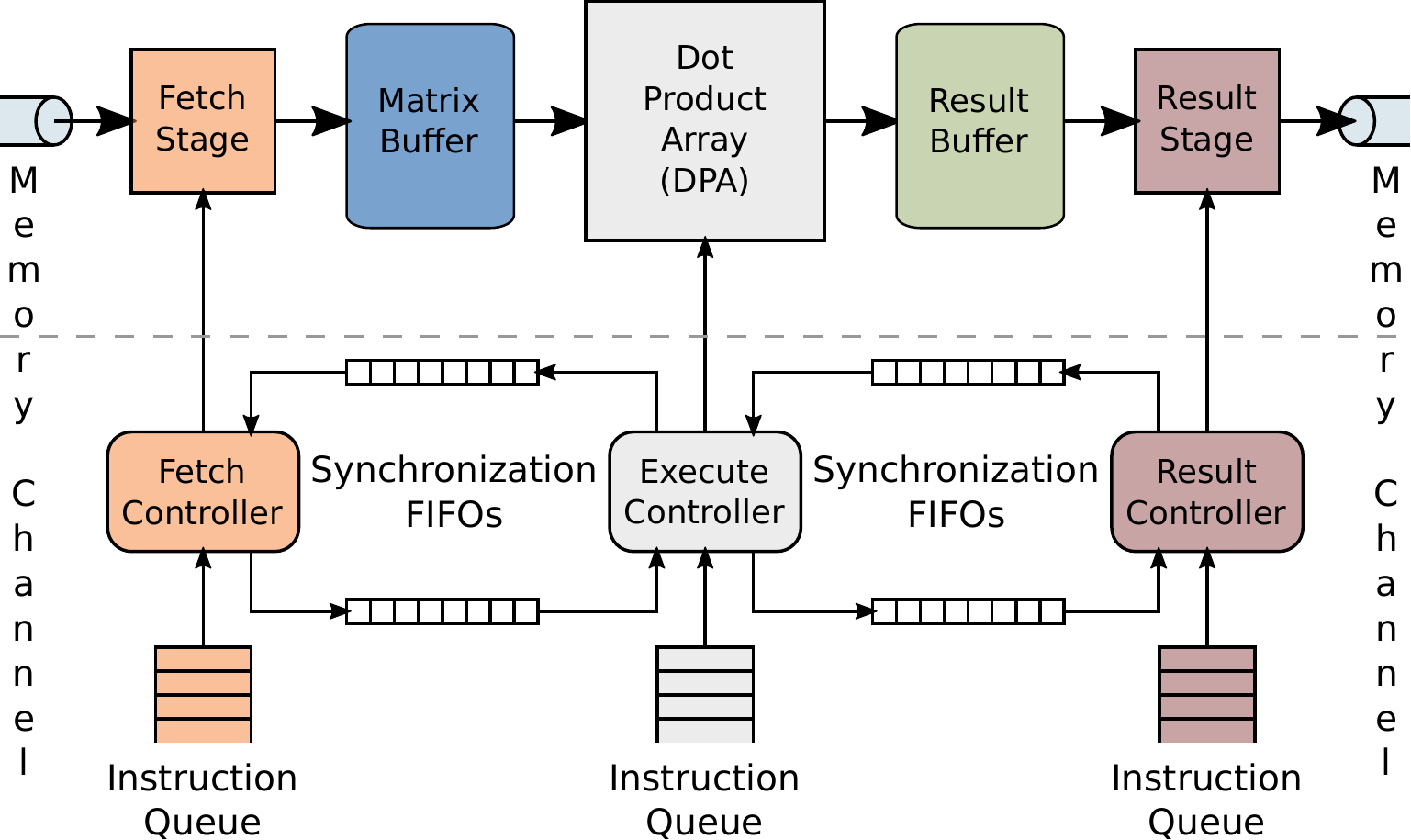}
  \vspace{-1mm}
  \caption{Overview of \OurScheme{}'s' hardware architecture.}
  \label{fig:hw_overview}
  \vspace{-3mm}
\end{figure}

\autoref{fig:hw_overview} provides an overview of the \OurScheme{}
hardware. %
The architecture is organized into three pipeline stages \textit{fetch},
\textit{execute}, and \textit{result}. %
Each stage communicates data to the next stage via shared on-chip memory
buffers. %
Inter-stage synchronization is achieved by blocking reads and writes to
synchronization FIFOs. %
All stage operations, including datapath control and synchronization, are
controlled by instructions, which are fetched from instruction queues and
executed in order. %


\begin{figure}[t]
  \centering
  \includegraphics[width=1\columnwidth]{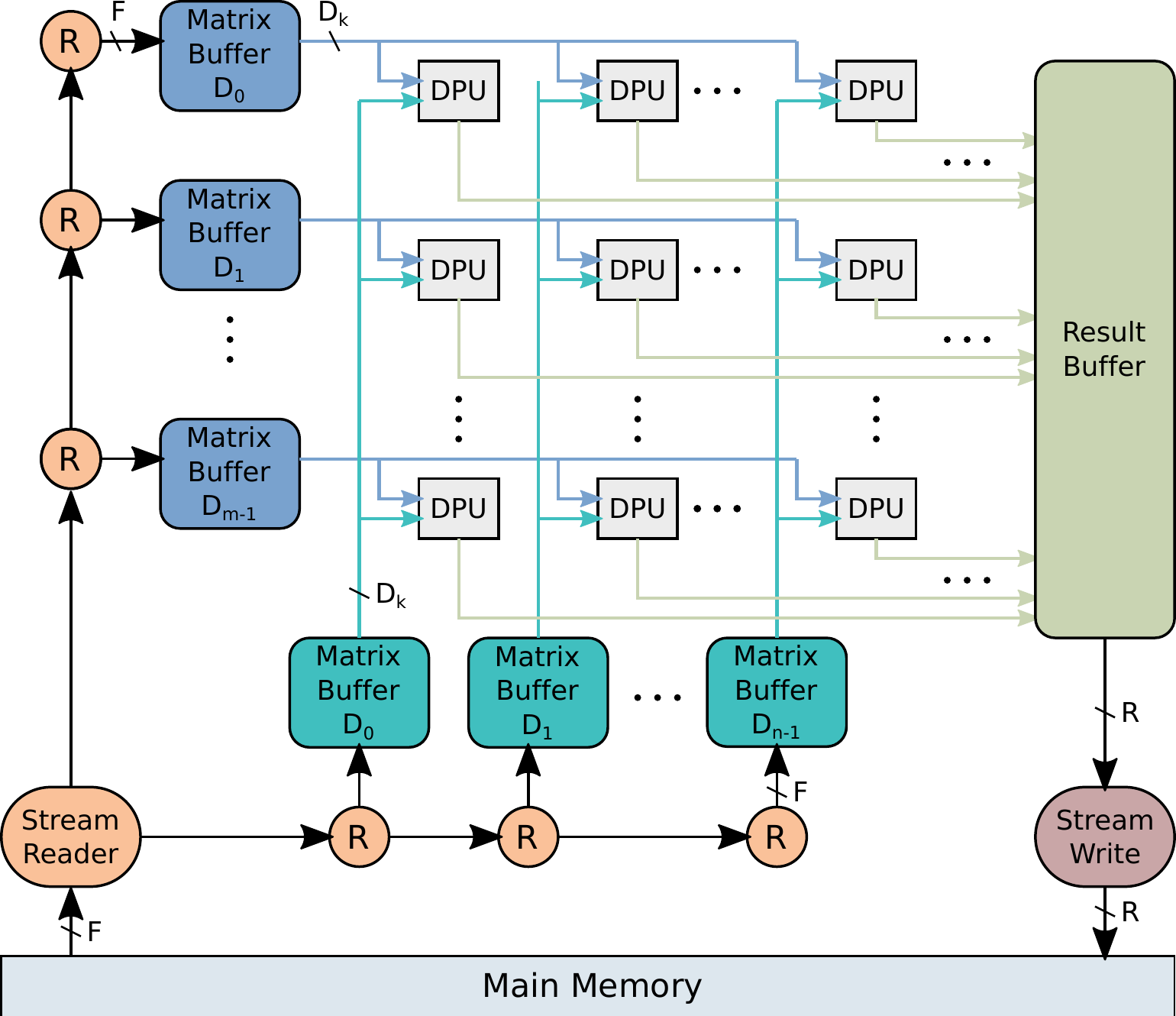}
  \caption{Key components of the \OurScheme{} datapath.}
  \label{fig:hw_datapath}
\end{figure}

The core of the hardware architecture is the bit-serial matrix-matrix
multiplication datapath illustrated in \autoref{fig:hw_datapath}. %
Accelerator performance and resource usage can be controlled by the
parameters specified in \autoref{tab:parameters}. %

\begin{table}
  \caption{Key \OurScheme{} hardware parameters.}
  \label{tab:parameters}
  \centering
  \begin{tabular}{cl}
    \toprule Symbol & Description \\
    \midrule
    $D_m, D_n$ & Number of DPUs in the DPA\\
    $D_k$ & DPU input bit width (popcount width) \\
    $B_m, B_n$ & Depth of input matrix buffers \\
    $B_r$ & Depth of result matrix buffer \\
    $A$ & Accumulator bitwidth \\
    $F$ & Main memory read channel bit width \\
    $R$ & Main memory write channel bit width \\
    \bottomrule
  \end{tabular}
\end{table}

\subsubsection{The Fetch Stage}
\label{sec:fetch_stage}

is responsible for reading matrix data from main memory and populating the
matrix buffers with data. %
Internally, the fetch stage contains a simple DMA engine and route generator
called a \textit{StreamReader}, as well as a linear array interconnect. %
The StreamReader sends read requests to main memory and determines where read
responses are to be written, as specified by fetch instructions. %
The read data and its destination form a packet that is carried through the
interconnect to the appropriate matrix buffer. %
The interconnect is bandwidth-matched to the main-memory read channel to
avoid any bottlenecks and ensure efficient use of off-chip bandwidth. %
The synchronization with the execute stage is ensured prior to fetching data,
which greatly simplifies the design of the network as there is no
backpressure. %
The fetch stage can be scaled at design time to match the memory read
bandwidth ($F$) of a particular platform. %


\subsubsection{The Execute Stage}
\label{sec:execute_stage}

is responsible for performing the matrix multiplication on the data present in
the matrix buffers. %
The core of the stage consists of an array of dot product units (DPUs), where
each DPU is fed with a design-time configurable number of bits ($D_k$) from the
left-hand-side and right-hand-side matrix buffers. %
The DPUs on the same row of the data processing array is fed with the same data
broadcasted by the left-hand-side matrix buffer. %
Similarly, the DPUs on the same column is fed with the same data broadcasted by
the right-hand-side matrix buffer (\autoref{fig:hw_datapath}). %
A single software controllable sequence generator is responsible for reading out
the appropriate data from the matrix buffers. %
The same generated sequence is used for both the left- and right-hand-side
matrix buffers but with different offsets. %
The execute stage can easily be scaled at design time by configuring the number
of rows ($D_M$) and columns ($D_N$) of DPUs. 

\begin{figure}[t]
  \centering
  \includegraphics[width=1\columnwidth]{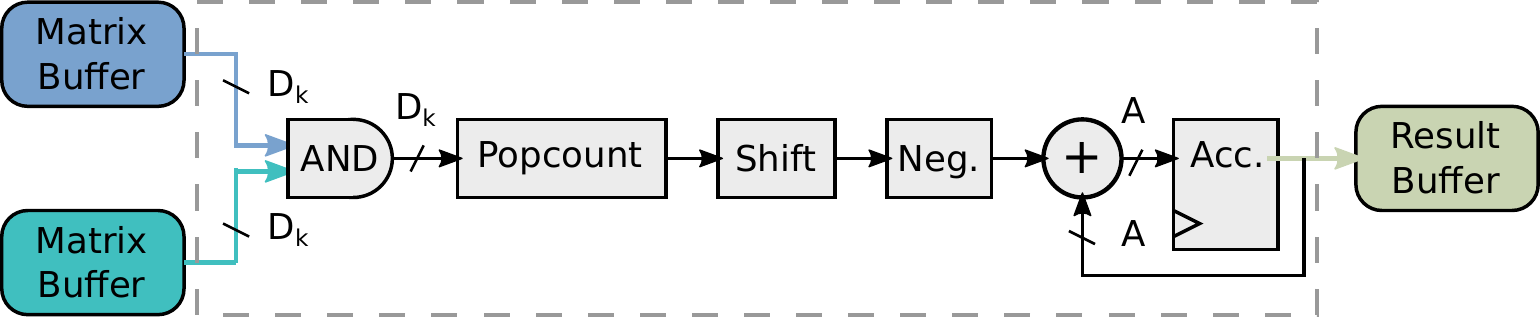}
  \caption{The \OurScheme{} dot product unit (DPU).}
  \label{fig:hw_dpu}
\end{figure}

The DPU pipeline can be seen in \autoref{fig:hw_dpu}. %
The DPU computes a partial result of the dot product between a row and
column of two bit-matrices, line 13 in
Algorithm~\ref{alg:bit-serial_matrix_multiplication}. %
The single-bit multiplications are performed by a multi-bit logic
\textsc{And} operation and the summation is a simple population count
(popcount) of the result. %
The weight in Algorithm~\ref{alg:bit-serial_matrix_multiplication} is
implemented by a left-shift unit and optional negation, which are
controllable by software. %
The partial results are accumulated and stored in a register (Acc.) of
width $A$, which is typically 32 bits
\cite{umuroglu_jahre:CASES2017, umuroglu+:FPGA2017finn} to avoid overflow. %

\subsubsection{The Result Stage}
\label{sec:result_stage}

is responsible for writing the results generated by the execute stage to main
memory. %
The stage consists of a \textit{StreamWriter}, which contains a downsizer
(wide-in-narrow-out) to resize the array of results into the appropriate width
needed by the memory channel and a DMA engine with striding support to carry
out the actual memory write operations. %
The striding is needed to produce the result matrix one tile at a time. %
When the execute stage has produced a new set of results, the accumulated
dot-products are written to the result buffer from which the result stage writes
them to main memory. %
This enables the two stages to work independently and to overlap computations
and data transfers. %
The result stage can be scaled at design time to match the memory write
bandwidth ($R$) of a particular platform. %


\subsection{Cost Model}
\label{sec:cost_model}

For any parametrizable overlay architecture, it is beneficial to provide a model
of how the FPGA resource usage relates to configuration parameters. %
This enables quick performance estimation when scaling to larger devices. %

\subsubsection{LUT cost}
\label{sec:costmodel_LUT}
We propose the following equations to model the LUT usage of a \OurScheme{}
instance:
\newcommand{\lutcost}[1] {\mathrm{LUT}_{\mathrm{#1}}}
\newcommand{\mconst}[1] {\alpha_{\mathrm{#1}}}
\newcommand{\aconst}[1] {\beta_{\mathrm{#1}}}
\newcommand{\bramcost}[1] {\mathrm{BRAM}_{\mathrm{#1}}}

\begin{subequations}
\vspace{-2mm}
\begin{equation} \label{eqn:lutcost_total}
    \lutcost{total} = \lutcost{base} + \lutcost{array}
\end{equation}
\vspace{-5mm}
\begin{equation} \label{eqn:lutcost_array}
  \lutcost{array} = D_m \cdot D_n \cdot (\lutcost{DPU} + \lutcost{res})
\end{equation}
\vspace{-5mm}
\begin{equation} \label{eqn:lutcost_dpu}
  \lutcost{DPU} = \mconst{DPU} \cdot D_k + \aconst{DPU}
\end{equation}
\end{subequations}

\autoref{eqn:lutcost_total} breaks the total cost into $\lutcost{base}$,
which covers the DPA size-independent LUT usage such as the DMA engines
and other fixed platform infrastructure, and $\lutcost{array}$ which covers
the DPA size-dependent part.
In turn, \autoref{eqn:lutcost_array} further breaks down $\lutcost{array}$
into LUT cost for the DPU and for result generation, multiplied by the array
size.
Finally, we model $\lutcost{DPU}$ as a linear function of the popcount
width $D_k$ in \autoref{eqn:lutcost_dpu}, and $\lutcost{res}$ as a constant.
The constants $\mconst{DPU}, \aconst{DPU}, \lutcost{base}$ and $\lutcost{res}$ are
determined empirically in \autoref{sec:synthesis}.

\subsubsection{BRAM cost}
\label{sec:costmodel_BRAM}
Assuming dual-port $36 \cdot 1024$-bit Xilinx BRAMs, we model the BRAM usage as
follows: %

\begin{subequations}
  \begin{equation} \label{eqn:bramcost_total}
      \bramcost{total} = \bramcost{base} + \bramcost{array}
  \end{equation}
\begin{equation} \label{eqn:bramcost_array}
    \bramcost{array} = \ceil*{\frac{D_k}{32}}  \cdot \left (D_m \cdot \ceil*{\frac{B_m}{1024}} + D_n \cdot \ceil*{\frac{B_n}{1024}} \right )
\end{equation}
\end{subequations}

In \autoref{eqn:bramcost_total}, $\bramcost{base}$ refers to the BRAMs used for
DPA-size independent infrastructure, such as DMA buffers and instruction
queues. %
$\bramcost{array}$ is the cost for the input matrix buffers. %
We use 32 of the native 36-bit width due to constraints from the fetch stage,
since DRAM buses are typically power-of-two-wide and we require BRAM read/write
widths to be an integer multiple of each other. %
We assume that the result matrix buffer consists of small LUTRAM buffers, and cover
their cost in \autoref{eqn:lutcost_array}. %

\subsection{Programming \OurScheme{}}
\label{sec:sw_stack}
\newcommand{\ISA}{Instruction}
\newcommand{\channelID}{Channel ID}
\newcommand{\schannelID}{channel ID}
\newcommand{\PutToken}{\texttt{Signal}}
\newcommand{\GetToken}{\texttt{Wait}}
\newcommand{\RunInst}{\texttt{Run}}
\newcommand{\RunFetch}{\texttt{RunFetch}}
\newcommand{\RunExecute}{\texttt{RunExecute}}
\newcommand{\RunResult}{\texttt{RunResult}}
\newcommand{\putverb}{-}
\newcommand{\getverb}{-}



\OurScheme{} provides programmability through the use of instructions that
control each of the pipeline stages. %
Taking into account the dimensions of the input matrices and the data layout in
memory it is possible for a programmer to perform scheduling in various ways. %
The capabilities facilitated by these instructions and their usage are
illustrated in this section. %

\subsubsection{\ISA{}s}

There are three types of instructions per pipeline stage in \OurScheme{}, namely
\GetToken{}, \PutToken{} and \RunInst{}. %
\autoref{tab:ins_desc} provides a summary of these instructions with the usage
described as follows: %

\begin{table}
  \caption{\OurScheme{}'s Instruction Summary}
  \label{tab:ins_desc}
  \centering
\begin{tabular}{ll}
\hline
Instruction type & Fields \\
\hline
\GetToken{} \& \PutToken{} & Associated FIFO: \\
 & ~~ Fetch stage: Execute \\
 & ~~ Execute stage: Fetch or Result \\
 & ~~ Result stage: Execute \\
\hline
\RunFetch{} & Source (main memory) parameters: \\
 & ~~ Base address \\
 & ~~ Block size (bytes) \\
 & ~~ Block offset (bytes) \\
 & ~~ Number of blocks to fetch \\
 & Destination (matrix buffer) parameters: \\
 & ~~ Matrix buffer offset \\
 & ~~ Starting matrix buffer \\
 & ~~ Range of matrix buffers \\
 & ~~ Consecutive words per matrix buffer \\
\hline
\RunExecute{} & Matrix buffer offset \\
 & Weight \\
 & Accumulator reset \\
\hline
\RunResult{} & Result base address in main memory \\
 & Address offset \\
\hline
\end{tabular}
\end{table}

\paragraph{The Synchronization Instructions}
are used for synchronization between two different pipeline stages. %
The \PutToken{} instruction issues a token to the associated synchronization
FIFO, while the \GetToken{} instruction blocks on the associated synchronization
FIFO until it receives a token. %
For both the fetch and result stage the only associated synchronization FIFO is
their respective FIFO for the execute stage. %
The execute stage has consequently two associated FIFOs for synchronization with
either the fetch or the result stage. %
The tokens do not convey any information and a programmer is free to decide what
each synchronization represents, e.g., that a particular matrix buffer is now
full or empty. %



\paragraph{The Run Instructions}
are used to carry out the particular function of a pipeline stage. %

The \RunFetch{} instruction specifies from where in main memory to read data and
the destination matrix buffers to store read data. %
The parameters with regard to main memory are: %
\textit{i)} the base address from where the fetch should begin, %
\textit{ii)} the size of the contiguous block to be fetched, %
\textit{iii)} the offset between such blocks (providing strided accesses), and %
\textit{(iv)} the number of blocks to be fetched. %
The parameters with regard to matrix buffers are: %
\textit{i)} the buffer offset at which to start writing data, %
\textit{ii)} the matrix buffer to begin writing to (all buffers are enumerated
from zero to $D_m \cdot D_n - 1$), %
\textit{iii)} the range of matrix buffers to be written (number of consecutive
buffers), and %
\textit{iv)} the number of consecutive words to be written in each matrix buffer
before switching to the next. %
These set of parameters enable consecutive data blocks to be placed in one
matrix buffer before moving to the next or to place the blocks in a cyclic
fashion across a range of buffers. %


The \RunExecute{} instruction specifies %
the matrix buffer offset from where to begin reading data, %
the weight controlling the shift amount and if the dot product should be negated
(line 8 in Algorithm~\ref{alg:bit-serial_matrix_multiplication}), and %
the possibility to reset the accumulators before performing any computations. %


The \RunResult{} instruction specifies the base address of the result matrix
stored in main memory and an offset to which the current results are to be
written. %



\newcommand{\mult}{$\cdot$}
\subsubsection{Instruction Scheduling}
\label{sec:schedexample}
%
The \OurScheme{} instructions enable the possibility to tailor the
computation to the input matrix characteristics, e.g., by
taking their dimensions into account. %

\begin{figure}[t]
  \centering
  \includegraphics[width=1\columnwidth]{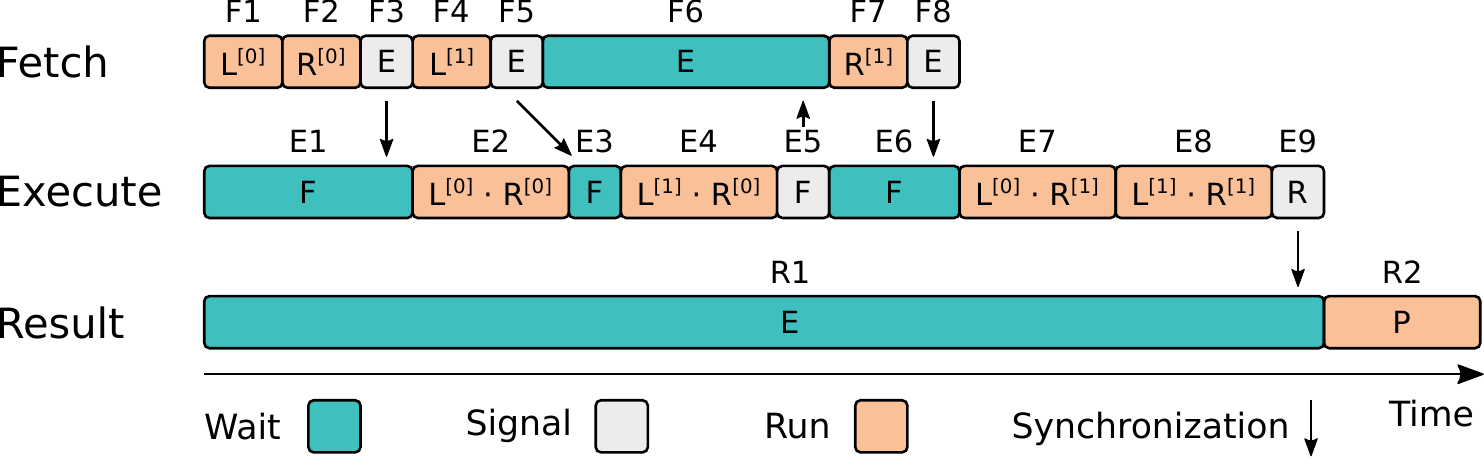}
  \caption{Timeline of the example schedule shown in \autoref{tab:instruction_queues}.}
  \label{fig:timeline}
\end{figure}

\autoref{fig:timeline} shows one possible schedule for the matrix multiplication
example in \autoref{fig:bit-serial_example}. %
Here, the DPA is assumed to be as large as the input matrices for simplicity. %
The computation would otherwise have to be divided into separate tiles resulting
in many more instructions. %
Furthermore, it is assumed that only three of the four binary matrices
($L^{[1]}$, $L^{[0]}$, $R^{[1]}$, and $R^{[0]}$) fit at the same time in the
matrix buffers to make the schedule slightly more interesting. %
The corresponding instructions for each pipeline stage can be seen in
\autoref{tab:instruction_queues}, with $P$ denoting the matrix that accumulates
the result of these operations. %

The fetch stage begins by fetching $L^{[0]}$ and $R^{[0]}$ (instruction F1 and
F2) and then signals the execute stage (F3) that it can perform the first
binary-matrix multiplication (E2). %
While the execute stage computes the dot product between $L^{[0]}$ and
$R^{[0]}$, the fetch stage continues fetching $L^{[1]}$, effectively achieving
an overlap between data fetch and execution (F4 and E2 performed in parallel). %
Once the execute stage finishes the first binary-matrix multiplication, it
receives the signal from the fetch stage (F5) that $L^{[1]}$ resides in the
matrix buffers (E3). %
The execute stage continues by executing $L^{[1]} \cdot R^{[0]}$ (E4) while the
fetch stage has to wait since all the buffer space is occupied (F6). %
When the execute stage finishes the matrix multiplication, it signals the fetch
stage (E5). %
Since $R^{[0]}$ is no longer needed, the fetch stage fetches $R^{[1]}$ (F7)
enabling the execute stage to finish the remaining matrix multiplications (E7
and E8). %
Once the execute stage has finished all binary matrix multiplications, it
signals the results stage (E9) which writes the result $P$ to main memory (R2). %

The schedule in \autoref{fig:timeline} causes the fetch stage and execute stage
to stall (F6 and E6) since there is not enough space to fetch $R^{[1]}$ before
$L^{[1]} \cdot R^{[0]}$ has been computed. %
An alternative schedule could be to split the binary matrices into tiles
enabling greater flexibility in what data to bring into the matrix buffers and
the possibility of overlapping fetch and execute. %

\begin{table}
  \caption{Initialized Instruction Queues for the Example Shown in \autoref{fig:bit-serial_example}}
  \label{tab:instruction_queues}
  \centering
  \begin{tabular}{l@{\hspace{0.3cm}}l@{\hspace{0.3cm}}l}
    \toprule
    Fetch & Execute & Result\\
    \midrule
F1 \RunInst{} L\(^{\text{[0]}}\) & E1 \GetToken{}  Fetch & R1 \GetToken{}  Execute\\
F2 \RunInst{} R\(^{\text{[0]}}\) & E2  \RunInst{} P =+  L\(^{\text{[0]}}\)
                                  \mult{}  R\(^{\text{[0]}}\) & R2
                                                                \RunInst{} P \\
F3  \PutToken{}  Execute & E3  \GetToken{}  Fetch & \\
F4  \RunInst{} L\(^{\text{[1]}}\) & E4  \RunInst{} P =+ L\(^{\text{[2]}}\)\mult{}  R\(^{\text{[0]}}\) & \\
F5  \PutToken{}  Execute & E5  \PutToken{}  Fetch & \\
F6    \GetToken{}  Execute & E6  \GetToken{}  Fetch & \\
F7 \RunInst{} R\(^{\text{[1]}}\) & E7  \RunInst{} P =+ L\(^{\text{[0]}}\)\mult{} R\(^{\text{[1]}}\) & \\
F8 \PutToken{}  Execute & E8  \RunInst{} P =+ L\(^{\text{[1]}}\)\mult{}  R\(^{\text{[1]}}\) & \\
 & E9 \PutToken{}  Result & \\
\bottomrule
  \end{tabular}
\end{table}

%% file: content/4-results.tex
\section{Evaluation}

\vspace{-1mm} 
We implement the \OurScheme{} parametrizable hardware generator in
Chisel~\cite{bachrach2012chisel} and use Xilinx Vivado 2017.4 for synthesis,
placement, and routing. %
We add registers to critical paths on the pipeline and enable register retiming
instead of manual floorplanning and timing optimizations to achieve higher clock
frequencies. %
We target the PYNQ-Z1 board, which has a Xilinx Z7020 FPGA with 53,200 LUTs, 140
BRAMs, and 3.2 GB/s of DRAM bandwidth. %


\vspace{-1mm} 
As binary operations are the building block for bit serial computations, we use
them as the common denominator for performance measurements. %
We treat \textsc{And} and popcount as analogues to multiplication and addition
when counting binary operations, i.e., a binary dot product between two
$n$-element binary vectors is counted as $2n$~binary operations. %


\input{content/4a-synth-results}
\input{content/4b-runtime-results}


%% file: content/4a-synth-results.tex
\subsection{Synthesis Results and Resource Cost}
\label{sec:synthesis}

\vspace{-1mm} 
We start by presenting synthesis results across a range of parameters for
different components of the \OurScheme{} architecture.
Our aim is to explore the resource cost of scaling performance along different
axes of parallelism and building up a hardware cost model in the process.
All data in this section is obtained by using out-of-context synthesis for the
Z7020, with a target clock period of 1~ns to prioritize timing optimizations.

\vspace{-1mm}
\pgfplotstableread[col sep=tab]{data/popcount.txt}\popcountdata
\begin{figure}[ht]
	\centering
	\resizebox{0.98\linewidth}{!}{
		\begin{tikzpicture}
			\begin{axis}[ylabel=$F_{\mathrm{max}}$~(MHz), axis y line*=right, ymin=0,
				width=\linewidth, height=4cm]
				\addplot [color=blue,mark=+,smooth] table [x=PopCWidth, y=fmax] {\popcountdata};
			\end{axis}

			\begin{axis}[xlabel=Popcount width $D_k$ (bits),ylabel=LUT,
				legend style={at={(0.5, 1.1)},anchor=south},
				legend columns=3,
				width=\linewidth, height=4cm]
				\addlegendimage{blue, only marks}
				\addlegendentry{$F_{\mathrm{max}}$}
				\addplot [color=red,only marks] table [x=PopCWidth, y=LUT] {\popcountdata};
				\addlegendentry{LUT}
				\addplot [red] table[y={create col/linear regression={y=LUT}}] {\popcountdata};
				\addlegendentry{%
				$\mathrm{LUT}=\pgfmathprintnumber{\pgfplotstableregressiona} \cdot D_k
				\pgfmathprintnumber[print sign]{\pgfplotstableregressionb}$ } %
			\end{axis}
		\end{tikzpicture}
	}
        \vspace{-2mm}
	\caption{Popcount unit LUT usage characterization.}
	\label{fig:popcountLUT}
\end{figure}
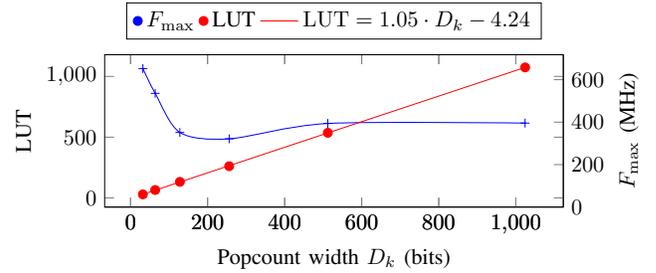

\subsubsection{Popcount}
\autoref{fig:popcountLUT} plots the LUT usage and maximum frequency
($F_{\mathrm{max}}$) versus input bitwidth for the popcount unit. %
We observe that the least squares regression line is a good fit for the LUT
usage, indicating a resource cost of approximately one LUT per input bit. %
This may be further improved by incorporating better compressor synthesis
techniques, as proposed by Preu{\ss}er~\cite{preusser2017generic}. %
The reported maximum clock frequency $F_{\mathrm{max}}$ is between 320 and
650~MHz for all tested popcount widths. %

\subsubsection{Dot Product Unit}
\begin{figure}
	\centering
	\resizebox{\linewidth}{!}{
		\begin{tikzpicture}
			\pgfplotstableread[col sep=tab]{data/dpu.txt}\dpudata;
			\begin{axis}[ylabel=Cost (LUT/bin.op.), axis y line*=right, ymin=0,
				width=\linewidth, height=4cm]
				\addplot [color=blue, smooth, mark=*] table [x=PopCWidth, y=LUT/op/cycle] {\dpudata};
				\label{luteff}
			\end{axis}
			\begin{axis}[xlabel=DPU width $D_k$ (bits),ylabel=Usage (LUT),
				legend style={at={(0.5, 1.1)},anchor=south},
				legend columns=3,
				width=\linewidth, height=4cm]
				\addlegendimage{blue, only marks}
				\addlegendentry{Op Cost}
				\addplot [color=red,only marks] table [x=PopCWidth, y=LUT] {\dpudata};
				\addlegendentry{Usage}
				\addplot [red] table[y={create col/linear regression={y=LUT}}] {\dpudata};
				\addlegendentry{%
				$\mathrm{LUT}=\pgfmathprintnumber{\pgfplotstableregressiona} \cdot D_k
				\pgfmathprintnumber[print sign]{\pgfplotstableregressionb}$ } %
			\end{axis}

		\end{tikzpicture}
	}
        \vspace{-5mm}
	\caption{DPU LUT usage and efficiency characterization.}
        \vspace{-1mm}
	\label{fig:dpuLUT}
\end{figure}
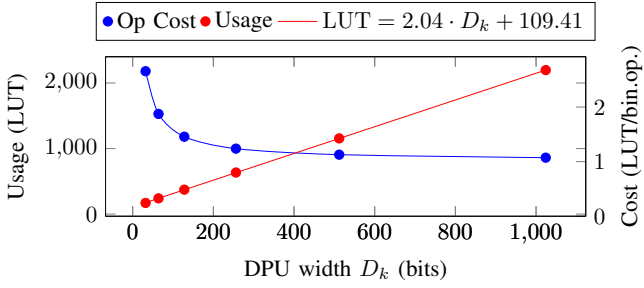

In addition to the popcount unit cost, the DPU cost includes the \textsc{And} operation,
barrel shifter, negator, and accumulator. %
We expect that the resource cost of the three latter gets amortized for larger
values of $D_k$ as their size grows proportional to $\mathrm{log}_2D_k$. %
\autoref{fig:dpuLUT} shows the  LUT usage as well as the LUT cost per binary
operation. 
We observe that the cost per operation starts at 2.8~LUTs for $D_k=32$ and
decreases to 1.07~LUTs for $D_k=1024$. 
The parameters $\mconst{DPU}$ and $\aconst{DPU}$ of the \OurScheme{} cost model
(\autoref{sec:costmodel_LUT}) are 2.04 and 109.41, respectively. %
For the tested bitwidths, the reported maximum frequency ($F_{\mathrm{max}}$) is
between 300 and 350~MHz. %

\subsubsection{Fetch and Result Stage}
We evaluate the cost of the fetch and result stages for a single 64-bit memory
channel on the PYNQ-Z1, with $F$=$R$=64, $A$=32, and $B_r$=2. %
The fetch stage includes a DMA engine and the interconnect to move data into
matrix buffers. %
We observe that the LUT cost of the fetch stage is approximated well by
$1.89 \cdot (D_m + D_n) + 463$. %
We do not include the $1.89 \cdot (D_m + D_n)$ component in the cost model since
it is small even for large DPAs. %
The result stage includes a DMA engine, result matrix buffers, and a downsizer
(parallel-to-serial unit), which are all implemented using LUTs. %
The result buffer requires approximately $87.3 \cdot D_m \cdot D_n$~LUTs, while
the DMA engine and the downsizer need $32.8 \cdot D_m \cdot D_n + 255$~LUTs. %
Completing the cost model, the fetch and result stages contribute
$463 + 255 = 718$~LUTs to $\lutcost{base}$, which may increase with more
advanced DMA engines, and the LUT cost per DPU associated with the result stage
is $\lutcost{res} = 87.3 + 32.8 = 120.1$. %
The DMA engine currently limits $F_{\mathrm{max}}$ to 200~MHz, and may be
pipelined to further increase $F_{\mathrm{max}}$ for the entire accelerator.

\subsubsection{Cost model validation}
We generated 34 different \OurScheme{} designs ranging from ($D_m$=2, $D_k$=64,
$D_n$=2) to ($D_m$=8, $D_k$=256, $D_n$=8) in size to validate the cost models
described in \autoref{sec:cost_model}. %
The BRAM predictions were 100\% accurate for this particular range of designs. %
\autoref{fig:model_accuracy} shows the LUT usage from synthesis results versus
the prediction from the cost model. %
The model's prediction is 93.8\% accurate on average. %
\autoref{fig:model_accuracy_vs_size} shows how the prediction error is affected by
the size of the design. %
We observe that large designs are accurately predicted, while smaller designs
tend to be overestimated by the model, likely due to the effect of additional
synthesis optimizations applied by Vivado for small designs. %

\pgfplotstableread[col sep=tab]{data/accel_pynq.txt}\accelpynqdata

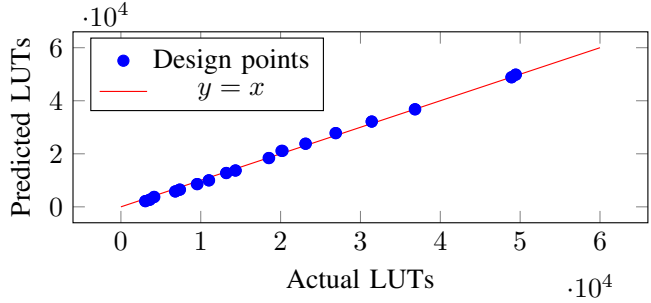
\begin{figure}
	\centering
	\resizebox{\linewidth}{!}{
		\begin{tikzpicture}
			\begin{axis}[xlabel=Actual LUTs ,ylabel=Predicted LUTs,
				legend pos=north west,
				width=\linewidth, height=4cm]
				\addplot [color=blue, only marks, mark=*] table [x=LUT, y=modelLUT] {\accelpynqdata};
				\addlegendentry{Design points}
				\addplot [domain=0:60000, color=red]{x};
				\addlegendentry{$y=x$}
			\end{axis}
		\end{tikzpicture}
	}
        \vspace{-5mm}
	\caption{Predicted vs actual LUT usage.}
	\label{fig:model_accuracy}
\end{figure}

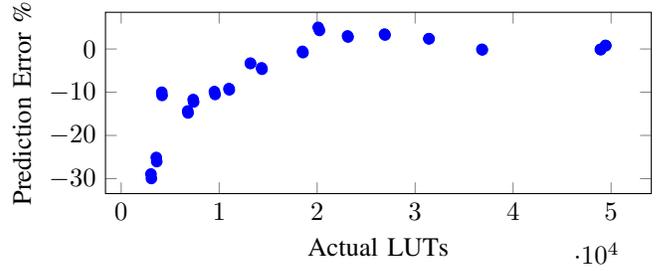
\begin{figure}
	\centering
	\resizebox{\linewidth}{!}{
		\begin{tikzpicture}
			\begin{axis}[xlabel=Actual LUTs,ylabel=Prediction Error \%,
				width=\linewidth, height=4cm]
				\addplot [color=blue, only marks, mark=*] table [x=LUT, y=errpct] {\accelpynqdata};
			\end{axis}
		\end{tikzpicture}
	}
        \vspace{-5mm}
	\caption{LUT cost model prediction error with design size.}
	\label{fig:model_accuracy_vs_size}
\end{figure}

\subsubsection{LUT-BRAM Tradeoffs}
\autoref{fig:lutbramtradeoffs} shows three \OurScheme{} instances with the same
performance and buffer depth but different overlay dimensions
($D_m$, $D_k$, $D_n$) and plot the number of BRAMs used and the LUT cost per
binary operation. %
We observe a tradeoff between BRAM and LUT cost by scaling different
parameters. %
We see that larger $D_k$ results in lower LUT cost, but requires more BRAMs to
deliver the bandwidth. %
Conversely, smaller $D_k$ needs fewer BRAMs, but has larger LUT cost. %
We note that the DPA dimensions should be matched to the workload dimensions for
higher efficiency, e.g., $D_n > 1$ is wasteful for matrix-vector multiplication,
but LUT and BRAM budget may impose additional constraints. %

\pgfplotstableread[col sep=tab]{data/accel_tradeoffs_constperf.txt}\acceldata
\begin{figure}
	\centering
	\resizebox{\linewidth}{!}{
		\begin{tikzpicture}
			\begin{axis}[xtick=\empty, ylabel=LUT/bin.op., axis y line*=right, ymin=0,
				width=\linewidth, height=4cm]
				\addplot [color=red, mark=+] table [x=cname, y=LUT/op] {\acceldata};
			\end{axis}
			\begin{axis}[xlabel=Configuration {($D_m$, $D_k$, $D_n$)}, ylabel=BRAM,
				legend style={at={(0.5, 1.1)},anchor=south},
				legend columns=3,
				xtick={1, 2, 3},
				xticklabels={{(2, 1024, 2)}, {(4, 256, 4)}, {(8, 64, 8)}},
				width=\linewidth, height=4cm]
				\addplot [color=blue, mark=*] table [x=cname, y=BRAM] {\acceldata};
				\addlegendentry{BRAM}
				\addlegendimage{red}
				\addlegendentry{LUT/bin.op.}
			\end{axis}
		\end{tikzpicture}
	}
	\caption{LUT vs BRAM tradeoffs for 1.6~binary~TOPS at 200~MHz.}
	\label{fig:lutbramtradeoffs}
\end{figure}
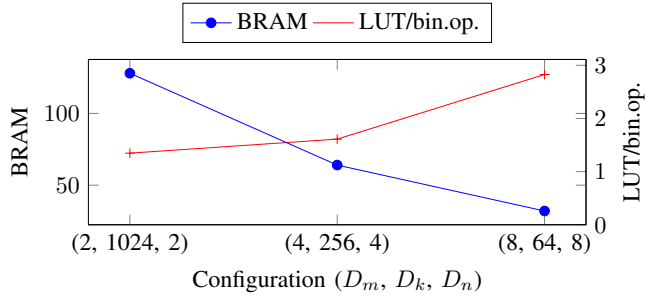

\begin{figure}
	\centering
	\resizebox{\linewidth}{!}{
		\begin{tikzpicture}
			\pgfplotstableread[col sep=tab]{data/multibitdpu.txt}\multibitdpudata;
			\begin{semilogxaxis}[xlabel=$w \times a$ dot products per cycle,ylabel=LUT/bin.op.,
				legend pos=south east, log basis x={2}, ymin=0,
				legend style={font=\tiny},
				legend columns=5,
				width=\linewidth, height=4cm, cycle list name=Dark2-8]
				\addplot [index of colormap=0 of Dark2-8, mark=*, smooth] table [x=SIMD, y=bismo] {\multibitdpudata};
				\addlegendentry{\OurScheme{}}
				\addplot [index of colormap=1 of Dark2-8, mark=*, smooth] table [x=SIMD, y=a2b1] {\multibitdpudata};
				\addlegendentry{$2 \times 1$}
				\addplot [index of colormap=2 of Dark2-8, mark=*, smooth] table [x=SIMD, y=a2b2] {\multibitdpudata};
				\addlegendentry{$2 \times 2$}
				\addplot [index of colormap=3 of Dark2-8, mark=*, smooth] table [x=SIMD, y=a3b2] {\multibitdpudata};
				\addlegendentry{$3 \times 2$}
				\addplot [index of colormap=4 of Dark2-8, mark=*, smooth] table [x=SIMD, y=a3b3] {\multibitdpudata};
				\addlegendentry{$3 \times 3$}
			\end{semilogxaxis}

		\end{tikzpicture}
	}
	\caption{Comparing the LUT/bin.op. cost of bit-serial and bit-parallel DPUs.}
	\label{fig:dpumultibitLUT}
\end{figure}
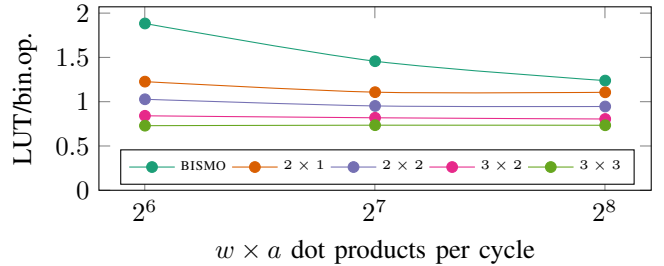

\subsubsection{Hardware Cost of Flexible Precision}
When required precision is known beforehand, a matrix multiplier that uses
fixed-precision bit-parallel arithmetic is the commonly used alternative, though
bit-serial could still be used. %
To quantify the overhead associated with bit-serial for those cases, we
implemented a version of the DPU with $w \times a$-bit multipliers instead of
\textsc{And}, an adder tree instead of popcount, and no shifter and negator. %
This bit-parallel DPU performs the equivalent of $2 \cdot w \cdot a \cdot D_k$
binary operations per cycle. %
\autoref{fig:dpumultibitLUT} compares the LUT cost for binary operation
equivalents between the \OurScheme{} DPU and several bit-parallel variants. %
We first observe that the LUTs per operation decreases with higher bit-parallel
precision, from 1.1 for $2 \times 1$ down to 0.73 for $3 \times 3$. %
Beyond $3 \times 3$ bits we did not observe further lowering of the LUT cost. %
As expected, bit-parallel DPUs have lower cost per bit operation compared to
bit-serial as they do not suffer from the shifter/negator overhead. %
For larger dot product sizes, the overhead is amortized and the worst-case gap
between \OurScheme{} and $3 \times 3$ closes down to 0.5 LUT per operation. %
We note that this is not a fully fair comparison since 1) the \OurScheme{}
hardware supports significantly larger precisions, and 2) our implementation
is not fully optimized down to the LUT level. %


%% file: content/4b-runtime-results.tex
\vspace{-1mm}
\subsection{Runtime Performance}
\label{sec:performance}

\vspace{-1mm}
 In this section, we assess the runtime performance and energy efficiency
achievable by \OurScheme{} instances running on the PYNQ-Z1. %
We assume that the input matrices are stored in DRAM using a bit-packed
data layout~\cite{umuroglu_jahre:CASES2017}, and that one matrix is transposed.
We create matrix multiplication workloads with different dimensions and
bitwidths, manually build the corresponding instruction sequences, and run
the workloads on the enumerated \OurScheme{} instances listed in
\autoref{tab:runtime_hw_configs} to evaluate how the overlay size interacts
with workload size. %

\begin{table}
  \centering
  \caption{\OurScheme{} instances for runtime measurements.}
  \label{tab:runtime_hw_configs}
  \begin{tabular}{ccccccc}
    \toprule
    \# & $D_m$ & $D_k$ & $D_n$ & LUT & BRAM & GOPS \\
    \midrule
    1 & 8 & 64 & 8 & 19545 (37\%) & 121 (86\%) & 1638.4 \\
    2 & 8 & 128 & 8 & 27740 (52\%) & 129 (92\%) & 3276.8 \\
    3 & 8 & 256 & 8 & 45573 (86\%) & 129 (92\%) & 6553.6 \\
    4 & 4 & 256 & 4 & 13352 (25\%) & 129 (92\%) & 1638.4 \\
    5 & 8 & 256 & 4 & 24202 (45\%) & 129 (92\%) & 3276.8 \\
    6 & 4 & 512 & 4 & 21755 (41\%) & 129 (92\%) & 3276.8 \\
    \midrule
    \multicolumn{7}{c}{$F=R=64$ and $F_{\mathrm{clk}}=200~\mathrm{MHz}$ unless otherwise stated.} \\
    \bottomrule
  \end{tabular}
  \vspace{-1mm}
\end{table}

\vspace{-1mm}
\subsubsection{Peak Binary Compute}
\label{sec:peakcompute_bin}
We start by
measuring the maximum achievable binary matrix multiply performance dictated
purely by the execute stage. %
For this experiment, we assume the matrices have already been fetched into
on-chip memory and disregard the cost of result writing. %
\autoref{fig:execeff} plots the achieved performance for different number of
columns as a percentage of observed peak performance. %
We observe that the efficiency increases with more columns, and that
instances with larger $D_k$ require wider matrices than smaller $D_k$ ones to be
efficient. %
As an example, for a matrix with 8192 columns, instance \#3 reaches 64\%
efficiency, while instance \#1 achieves 89\%. %
The inefficiency for narrow matrices is due to the lack of work to fill the DPA
pipeline, e.g., the DPA pipeline may be 10-deep but each dot product is finished
in 6 cycles. %
This can be remedied by issuing more work to the DPA without waiting for the
previous execution to finish, or by decreasing the DPA pipeline depth. %
Wide matrices achieve close to 100\% of the peak performance for all instances.

\pgfplotstableread[col sep=tab]{data/execeff_64.txt}\execeffs
\pgfplotstableread[col sep=tab]{data/execeff_128.txt}\execeffm
\pgfplotstableread[col sep=tab]{data/execeff_256.txt}\execeffl
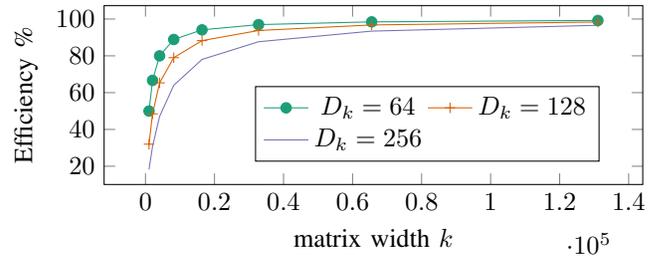
\begin{figure}
	\centering
	\resizebox{\linewidth}{!}{
		\begin{tikzpicture}
			\begin{axis}[xlabel=matrix width $k$,ylabel=Efficiency \%,
				legend style={at={(0.6, 0.55)},anchor=north},
				legend columns=2,
				width=\linewidth, height=4cm]
				\addplot[mark=*, index of colormap=0 of Dark2-8] table [x=K, y=ExecEff] {\execeffs};
				\addlegendentry{$D_k=64$}
        \addplot[mark=+, index of colormap=1 of Dark2-8] table [x=K, y=ExecEff] {\execeffm};
				\addlegendentry{$D_k=128$}
        \addplot[index of colormap=2 of Dark2-8] table [x=K, y=ExecEff] {\execeffl};
				\addlegendentry{$D_k=256$}
			\end{axis}
		\end{tikzpicture}
	}
	\caption{Execute stage efficiency depending on $D_k$ and matrix width $k$.}
	\label{fig:execeff}
\end{figure}

\subsubsection{Peak Bit-Serial Compute}
\label{sec:peakbitserial}
Per Algorithm~\ref{alg:bit-serial_matrix_multiplication}, if the runtime of a
binary ($1 \times 1$) matrix multiplication of a given size is $t$, we expect
the runtime of a $w \times a$-bit matrix multiplication of the same size to be
$w \cdot a \cdot t$. %
\autoref{fig:multibit} plots the performance for $8 \times 2048 \times 8$ and
$8 \times 16384 \times 8$ with increasing $w, a$ on instance~\#2. %
We observe slightly better performance than the projected $w \cdot a \cdot t$
since multiple dot products are accumulated together for the multi-bit case,
behaving like a longer dot product and increasing the execute stage efficiency
(\autoref{fig:execeff}). %

\pgfplotstableread[col sep=tab]{data/multibit_k128_K2048.txt}\multibitressmall
\pgfplotstableread[col sep=tab]{data/multibit_k128_K16384.txt}\multibitreslarge
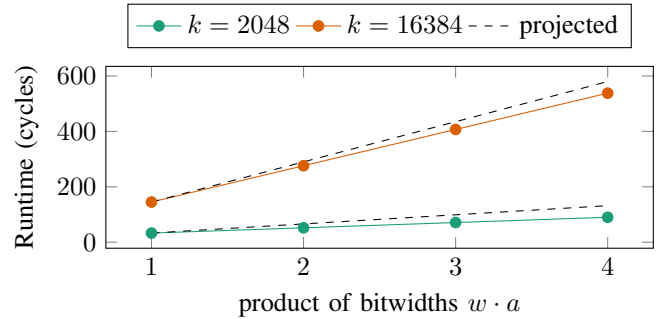
\begin{figure}
	\centering
	\resizebox{\linewidth}{!}{
		\begin{tikzpicture}
			\begin{axis}[xlabel=product of bitwidths $w \cdot a$, ylabel=Runtime (cycles),
        xtick=data,
				legend style={at={(0.5, 1.1)},anchor=south},
				legend columns=3,
				width=\linewidth, height=4cm]
				\addplot[mark=*, index of colormap=0 of Dark2-8] table [x=bitcomp, y=cycles] {\multibitressmall};
				\addlegendentry{$k=2048$}
        \addplot[mark=*, index of colormap=1 of Dark2-8] table [x=bitcomp, y=cycles] {\multibitreslarge};
				\addlegendentry{$k=16384$}
        \addplot[style=dashed] table [x=bitcomp, y=proj_bin] {\multibitreslarge};
        \addplot[style=dashed] table [x=bitcomp, y=proj_bin] {\multibitressmall};
        \addlegendentry{projected}
			\end{axis}
		\end{tikzpicture}
	}
	\caption{Runtime with increasing precision on instance \#2.}
	\label{fig:multibit}
\end{figure}


\begin{table}
\centering
  \caption{Power consumption data from \OurScheme{} instances on PYNQ-Z1.}
  \label{tab:power_results}
  \scriptsize
  \begin{tabular}{crrrrcc}
    \toprule
    Configuration & \multicolumn{4}{c}{Power (W)} & Binary & Binary \\
    \cline{2-5}
    $(\mathrm{Instance}, F_{\mathrm{clk}})$ & Idle & Exec & F \& R & Full & GOPS & GOPS/W \\
    \midrule
    (\#1, 200 MHz) & 2.53 & +0.33 & +1.09 & 4.07 & 1638 & 402.16 \\
    (\#2, 100 MHz) & 2.10 & +0.19 & +0.87 & 3.11 & 1638 & 527.51 \\
    (\#3, ~50 MHz) & 1.76 & +0.30 & +0.63 & 2.53 & 1638 &646.39 \\
    (\#4, 200 MHz) & 2.53 & +0.34 & +1.09 & 3.86 & 1638 &424.98 \\
    (\#5, 100 MHz) & 2.05 & +0.24 & +0.92 & 3.06 & 1638 &536.02 \\
    \midrule
    (\#3, 200 MHz) & 2.87 & +0.71 & +1.19 & 4.64 & 6554 &1413.39 \\
    \bottomrule
  \end{tabular}
\end{table}

\begin{table*}
  \caption{Comparing \OurScheme{} to recent work.}
  \vspace{-1mm}
  \label{tab:relatedwork}
  \centering
  \begin{tabular}{ccccrrc}
    \toprule
    Work & Platform & Type & Precision & Binary GOPS & GOPS/W & \\
    \midrule
    \OurScheme{} & Z7020 on PYNQ-Z1 & FPGA & bit-serial & 6554 & 1413.40 &
    \parbox[t]{2mm}{\multirow{6}{*}{\rotatebox[origin=c]{90}{incl. DRAM}}} \\
    FINN \cite{umuroglu+:FPGA2017finn} & Z7045 on ZC706 & FPGA & binary & 11613 & 407.50 & \\
    Moss et al. \cite{moss2018customizable} & GX1150 on HARPv2 & FPGA & reconfigurable & 41 & 849.38 & \\
    Umuroglu et al. \cite{umuroglu_jahre:CASES2017}$\dagger$ & Cortex-A57 on Jetson TX1 & CPU & bit-serial & 92 & 18.80 & \\
    Pedersoli et al. \cite{pedersoli2017espresso}$\dagger$ & GTX 960 & GPU & limited bit-serial & 90909 & 757.60 & \\
    Judd et al. \cite{judd2016stripes}$\dagger$ & ASIC & ASIC & limited bit-serial & 128450 & 4253.30 & \\
    \midrule
    \OurScheme{} & Z7020 on PYNQ-Z1 & FPGA & bit-serial & 6554 & 1889.70 & \parbox[t]{2mm}{\multirow{4}{*}{\rotatebox[origin=c]{90}{excl. DRAM}}}\\
    FINN \cite{umuroglu+:FPGA2017finn} & Z7045 on ZC706 & FPGA & binary & 11613 & 992.50 & \\
    Umuroglu et al. \cite{umuroglu_jahre:CASES2017}$\dagger$ & Cortex-A57 on Jetson TX1 & CPU & bit-serial & 92 & 43.80 & \\
    Umuroglu et al. \cite{umuroglu_jahre:CASES2017}$\dagger$ & i7-4790 & CPU & bit-serial & 355 & 12.20 & \\
    \multicolumn{7}{c}{\emph{$\dagger$ indicates our experiments from released code or projections based on paper.}} \\
    \bottomrule
  \end{tabular}
  \vspace{-2mm}
\end{table*}

\subsubsection{Stage Overlap}
We now quantify the performance gain by overlapping the fetch, execute and result
stages for larger matrix multiplications.
We create an instruction sequence to run a $256 \times 4096 \times 256$
binary matrix multiplication on instance \#1, similar to the example in
\autoref{sec:schedexample}.
The input matrices here are twice the size of the on-chip memory.
By overlapping the operation of different stages, the multiplication finishes
in 121133 cycles, achieving a speedup of $2.2\times$ compared to the 266510 cycles
when the stages are executing without overlap.

\subsubsection{Power Consumption}
\OurScheme{}'s power efficiency is measured using a PYNQ-Z1 board powered over a
USB port with a power meter attached while running one or more stages in a
loop. %
\autoref{tab:power_results} lists five instances where the frequency
($F_{\mathrm{clk}}$) is adjusted to achieve the same peak binary performance
(GOPS) followed by the top-performing \OurScheme{} instance. %
We list four power readings: %
the idle power with no stages running,
the increment from idle with only the execute stage running,
the increment with only the fetch and result stages running, and
the full power with all stages running. %
We find that on average the execute stage contributes 9.7\% of the full power
consumption, while the fetch and result stages contribute 27.2\% and the idle
power constitutes 65.6\%. %
For the cases with constant performance, we see that a large but slow-clocked
design achieves $1.5\times$ better power efficiency than a small but
fast-clocked design, similar to what is reported in
FINN~\cite{umuroglu+:FPGA2017finn}. %
The majority of this increase in power efficiency can be attributed to lower
idle power due to a slower clock.


%% file: content/5-relatedwork.tex
\vspace{-2mm}
\section{Related Work}
\label{sec:relatedwork}

\vspace{-2mm}
\autoref{tab:relatedwork} compares \OurScheme{} against several recently-proposed
implementations for low-precision matrix multiplication, 
using peak binary performance and performance per watt as metrics.
The top part of the table includes DRAM power, while the bottom part only
considers on-chip compute and memory power.
To our knowledge, \OurScheme{} is the first FPGA implementation for bit-serial
matrix multiplication, but comparable related work on binarized neural networks
by Umuroglu et al.~\cite{umuroglu+:FPGA2017finn} and low-precision matrix
multiplication by Moss et al.~\cite{moss2018customizable} report 
$3.5\times$ and $1.6\times$, respectively, lower power efficiency than ours.
Although the GPU binary matrix multiplication kernels proposed by Pedersoli et
al.~\cite{pedersoli2017espresso} achieve an impressive 90~TOPS for large
matrices, their work does not report power measurements.
Assuming a power consumption of 120~W for the GTX 960, \OurScheme{} achieves
$1.9\times$ better power efficiency in comparison.
On CPUs, the single-threaded implementation by Umuroglu and
Jahre~\cite{umuroglu_jahre:CASES2017} performed far worse than \OurScheme{},
and is still outperformed by more than an order of magnitude even assuming $4\times$
performance improvement with multi-core parallelization.
Finally, Stripes by Judd et al.~\cite{judd2016stripes} outperforms ours by $3\times$
due to the performance and efficiency of an ASIC~implementation.

%% file: content/6-conclusion.tex
\section{Conclusion}

We have presented \OurScheme{}, a bit-serial matrix multiplication overlay that
can scale its precision to match an application's computational requirements and
its hardware to match available system resources. %
The proposed cost model accurately predicts FPGA resource utilization and enables
quick performance estimations. %
\OurScheme{} is software programmable, providing the possibility to adapt its
execution to the dimension and precision of any input matrix. %
Our evaluation indicates that \OurScheme{} achieves a peak
performance of 6.5 TOPS with an energy efficiency of up to 1.4 TOPS/W on a
PYNQ-Z1 board. %





